\documentclass[10pt,twocolumn,letterpaper]{article}
\usepackage{authblk}

\usepackage[pagenumbers]{styleguide} 

\usepackage[dvipsnames]{xcolor}
\usepackage{ulem}

\usepackage{tikz}
\usepackage{pgfplots}
\usepackage{stfloats}
\usepackage{graphicx}
\usepackage{placeins}
\usepackage{array}
\newcolumntype{x}[1]{>{\centering\arraybackslash\hspace{0pt}}p{#1}}

\usepackage{amsmath, booktabs, caption, longtable}

\definecolor{cvprblue}{rgb}{0.21,0.49,0.74}

\usepackage[pagebackref,breaklinks,colorlinks,citecolor=cvprblue]{hyperref}

\usepackage{float}

\pgfplotsset{compat=1.18}

\usepackage{tabularx}

\title{SPINEPS – Automatic Whole Spine Segmentation of T2-weighted MR images using a Two-Phase Approach to Multi-class Semantic and Instance Segmentation.}

\begin{titlepage}
\end{titlepage}
\author[1, 2]{Hendrik Möller} 
\author[1, 2]{Robert Graf} 
\author[1]{Joachim Schmitt} 
\author[1]{Benjamin Keinert} 
\author[1, 2]{Matan Atad} 
\author[1, 3]{Anjany Sekuboyina}
\author[7]{Felix Streckenbach}
\author[1, 7]{Hanna Schön}
\author[1,2,5,6]{Florian Kofler} 
\author[9, 10]{Thomas Kroencke}
\author[9]{Stefanie Bette}
\author[11]{Stefan N Willich}
\author[11]{Thomas Keil}
\author[12]{Thoralf Niendorf}
\author[12]{Tobias Pischon}
\author[12]{Beate Endemann}
\author[3]{Bjoern Menze}
\author[2,4]{\\Daniel Rueckert}
\author[1]{Jan S. Kirschke}

\affil[1]{\footnotesize Department of Diagnostic and Interventional Neuroradiology, School of Medicine, Klinikum rechts der Isar, Technical University of Munich, Munich, Germany}
\affil[2]{\footnotesize Department for AI in Medicin, Klinikum Rechts Der Isar, Technical University of Munich, Munich, Germany.}
\affil[3]{\footnotesize Department of Quantitative Biomedicine, University of Zurich, Zurich, Switzerland}
\affil[4]{\footnotesize Visual Information Processing, Imperial College London, London, United Kingdom}
\affil[5]{\footnotesize Helmholtz AI, Helmholtz Munich, Neuherberg, Germany}
\affil[6]{\footnotesize TranslaTUM - Central Institute for Translational Cancer Research, Technical University of Munich, Munich, Germany}
\affil[7]{\footnotesize Department of Diagnostic and Interventional Radiology, Pediatric Radiology and Neuroradiology, University Medical Center Rostock, Rostock, Germany.}
\affil[8]{\footnotesize Institute for Prevention and Cancer Epidemiology, Faculty of Medicine and Medical Center, University of Freiburg, Freiburg, Germany}
\affil[9]{\footnotesize Department of Diagnostic and Interventional Radiology, University Hospital Augsburg, Augsburg, Germany}
\affil[10]{\footnotesize Centre for Advanced Analytics and Predictive Sciences, Augsburg University, Augsburg, Germany}
\affil[11]{\footnotesize Project Division Epidemiology and Prevention, Institut für Sozialmedizin, Epidemiologie und Gesundheitsökonomie, Charité - Universitätsmedizin Berlin, Germany}
\affil[12]{\footnotesize Max Delbrück Center for Molecular Medicine in the Helmholtz Association, Berlin, Germany}

\begin{document}

\maketitle
\vspace{-1cm}

\begin{abstract}

    \textbf{Purpose.} To present SPINEPS, an open-source deep learning approach for semantic and instance segmentation of 14 spinal structures (ten vertebra substructures, intervertebral discs, spinal cord, spinal canal, and sacrum) in whole body T2w MRI.
    
    \textbf{Methods.} During this HIPPA-compliant, retrospective study, we utilized the public SPIDER dataset (218 subjects, 63\% female) and a subset of the German National Cohort (1423 subjects, mean age 53, 49\% female) for training and evaluation. We combined CT and T2w segmentations to train models that segment 14 spinal structures in T2w sagittal scans both semantically and instance-wise. Performance evaluation metrics included Dice similarity coefficient, average symmetrical surface distance, panoptic quality, segmentation quality, and recognition quality. Statistical significance was assessed using the Wilcoxon signed-rank test. An in-house dataset was used to qualitatively evaluate out-of-distribution samples.

    \textbf{Results.} On the public dataset, our approach outperformed the baseline (instance-wise vertebra dice score 0.929 vs. 0.907, p-value$<0.001$). Training on auto-generated annotations and evaluating on manually corrected test data from the GNC yielded global dice scores of 0.900 for vertebrae, 0.960 for intervertebral discs, and 0.947 for the spinal canal. Incorporating the SPIDER dataset during training increased these scores to 0.920, 0.967, 0.958, respectively.
    
    \textbf{Conclusions.} The proposed segmentation approach offers robust segmentation of 14 spinal structures in T2w sagittal images, including the spinal cord, spinal canal, intervertebral discs, endplate, sacrum, and vertebrae. The approach yields both a semantic and instance mask as output, thus being easy to utilize. This marks the first publicly available algorithm1 for whole spine segmentation in sagittal T2w MR imaging.
\end{abstract}

\begin{table*}[!hbt]
    \centering
    \begin{tabular}{llll}
        \toprule
         & German National Cohort & SPIDER dataset & In-house dataset \\ \midrule
        Subjects & 1966 & 218 & 15 \\
        Modality & T2w sagittal scans & T1w and T2w sagittal scans & T2w sagittal scans \\
        Region & Cervical, Thoracic, Lumbar & Lumbar only & Cervical, Thoracic, Lumbar \\
        Sex (\% female) & 49 & 63 & 47 \\
        Mean age (yrs) $\pm$ SD & $52 \pm 11$ & N/A & $66 \pm 19$ \\
        \bottomrule
    \end{tabular}
    \caption{Demographics of the utilized cohorts. SD stands for standard deviation.}
    \label{tab:demographics}
\end{table*}

\begin{figure*}[!b]
    \centering
    \includegraphics[width=0.8\textwidth]{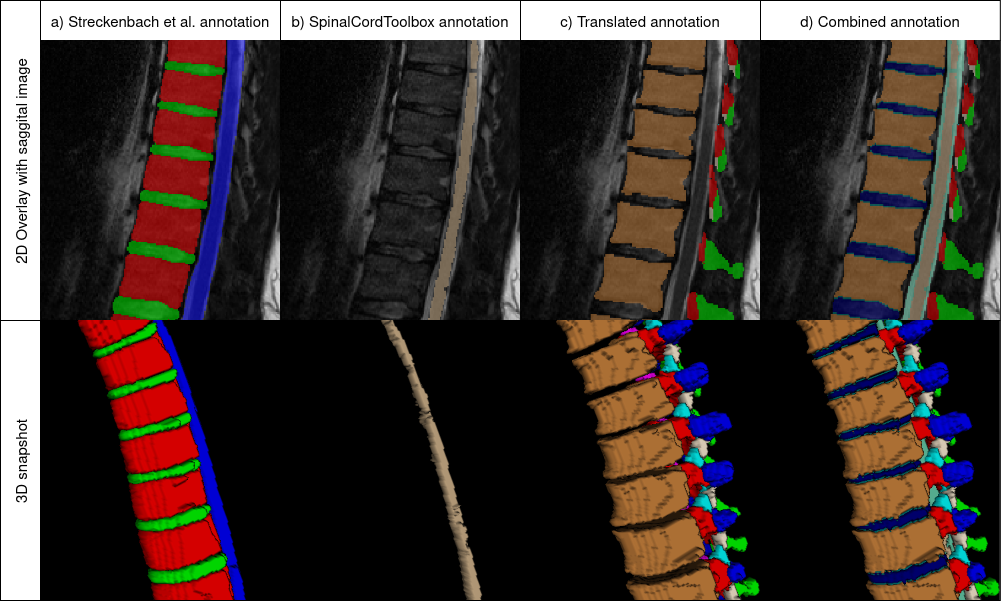}
    \caption{Showcase of the three automated annotations and its resulting combined annotation, as a 2D segmentation overlay and a 3D snapshot. a) shows the segmentation made with the training data from Streckenbach et al. \cite{streckenbach_application_2022}, b) the SpinalCordToolbox annotation, c) the annotations derived from translation, and d) is the combination of all three. We observed that the manual segmentation from Streckenbach et al. \cite{streckenbach_application_2022} is primarily block-shaped and incomplete, while the translated annotations often segmented too many voxels around the vertebra corpus.}
    \label{fig:combine}
\end{figure*}

\section{Introduction}
\label{sec:introduction}
MRI is commonly utilized to evaluate the spine in clinical practice, providing diagnostically useful data on intervertebral disc degeneration, endplate changes, and spinal canal/cord structure \cite{tamagawa_imaging_2022}. Due to the wealth of information contained in such imaging data and its non-invasive process, MRI is also acquired in epidemiologic studies with large sample sizes. One crucial part in the data analysis of such studies is the segmentation of relevant structures.

Segmentation is the task of assigning each voxel in an image a label, creating a segmentation mask. It is commonly categorized into semantic segmentation, where labels represent categories, and instance segmentation, where each label corresponds to a unique instance \cite{minaee_image_2022}. These segmentation masks are helpful in automatically quantifying image biomarkers and detecting and locating pathologies and abnormalities \cite{zhang_biomarker_2022}. Hence, they are often essential prerequisites for more extensive studies like the German National Cohort (NAKO) \cite{bamberg_whole-body_2015}.

Machine learning represents an established tool to solve the problem of semantic or instance segmentation \cite{yao_cnn_2023}. However, currently, there is no automatic approach for MRI images that segments the whole spine, including posterior elements like the spinous processes of vertebrae. In addition, most existing methods for T2w image segmentation are limited to the lumbar region and, therefore, not designed to segment the whole spine \cite{al_kafri_segmentation_2018, badarneh_semi-automated_2021, khalil_multi-scanner_2022, saenz-gamboa_automatic_2023, shi_efficient_2007, li_automatic_2021}. However, for CT imaging, instance segmentation is well established \cite{sekuboyina_verse_2021, payer_coarse_2023, chen_automatic_2015}. From existing models, we observed that Unet architectures \cite{navab_u-net_2015} struggle with instance segmentation, as the different instance labels belong to the same semantic structure, exemplarily described by Isensee et al. \cite{isensee_extending_2023}. This is unacceptable if such segmentation masks are used for statistical analysis, registration or medical intervention.
Finally, training a segmentation approach is laborious, as it usually requires manual annotations of the whole training data. To overcome the issue of manual annotations, Graf et al. \cite{graf_denoising_2023} successfully used translation to create artificial CT- from MR images. They used existing segmentation models for CT to create segmentation masks for MRIs and showed that this translation works well enough to transfer CT-level segmentation masks into MRIs.

The purpose of this paper is to (1) present SPINEPS, a two-phase approach to segment 14 spinal structures in the cervical, thoracic, and lumbar regions of T2w sagittal images both semantically and instance-wise, (2) to show how a combination of annotations derived from automated segmentation models and MR to CT image translation approaches can be utilized for training and, (3) make a pre-trained model publicly available to enable researchers to generate segmentation masks for their own datasets.

\begin{figure*}[!b]
    \centering
    \includegraphics[width=0.9\textwidth]{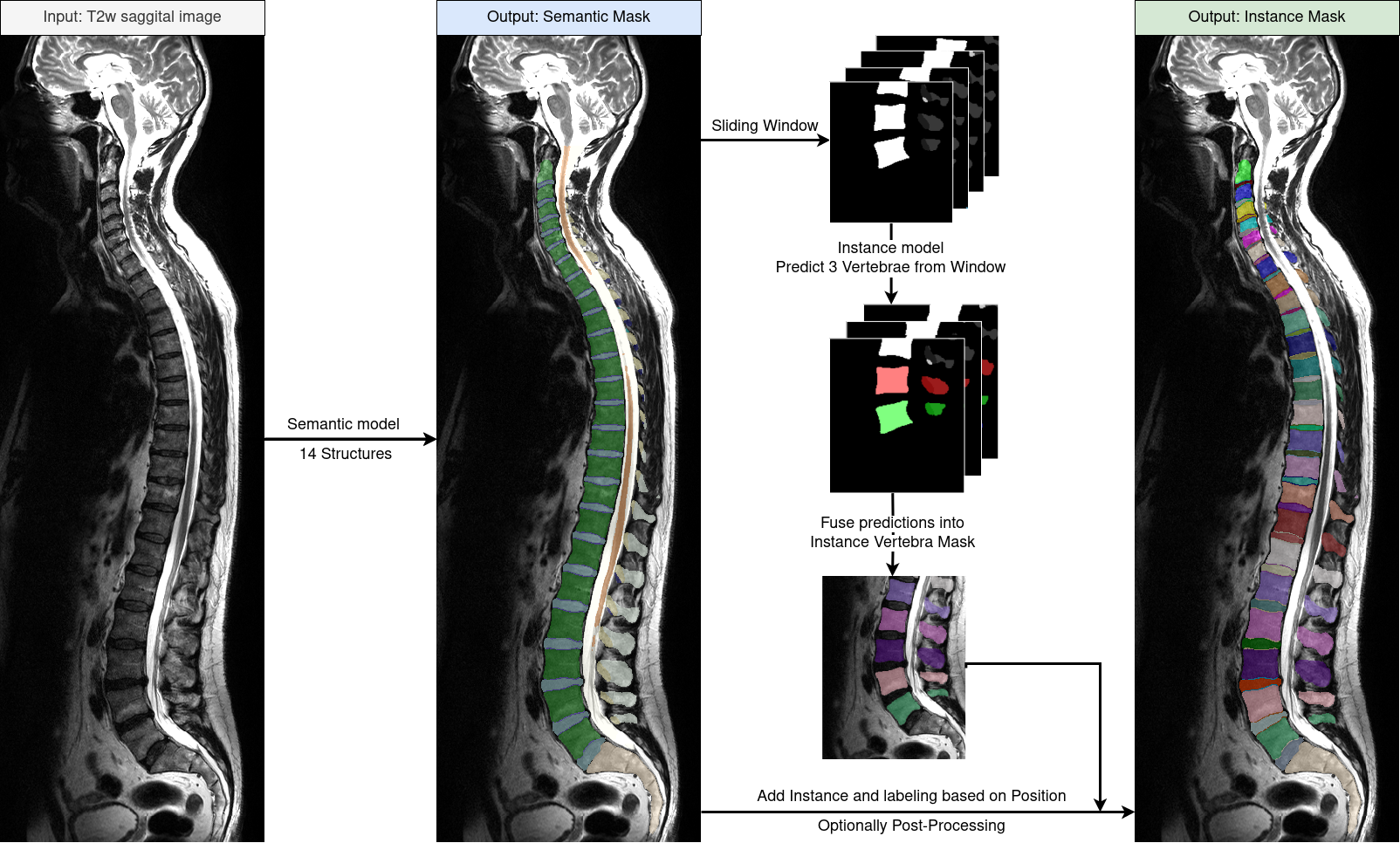}
    \caption{The data flow of our proposed method of inference on new T2w sagittal scans. The semantic model segments 14 different spine structures, regardless of field of view. Then, cutouts are made from that segmentation, which are fed into the instance model. The results are predictions for the individual vertebrae, which are fused together for the vertebra instance mask. Then, using the first segmentation, we zero out everything in the vertebra mask that is not present in the subregion mask. Finally,  we match intervertebral discs and endplate instances based on a center of mass analysis.}
    \label{fig:workflow}
\end{figure*}

\section{Methods}
\label{sec:methods}
This retrospective, HIPPA-compliant study utilized a public dataset, external dataset (participants gave informed consent) and in-house dataset (waived informed consent by the local ethics committee).

\subsection{Participants}
We utilized the public SPIDER dataset \cite{van_der_graaf_lumbar_2023}, a random German National Cohort (NAKO) subset and an in-house dataset (for demographics, see Table \ref{tab:demographics}). 18\% (39/218) random cases from SPIDER was used as test split. For a NAKO test split, eleven random subjects underwent manual correction by two experts with three and two years of experience, supervised by an expert with 22 years of experience. The in-house dataset was used for qualitative evaluation.

\subsection{Automated Annotations}
The manually annotated data from Streckenbach et al. \cite{streckenbach_application_2022} contains the vertebra corpus, IVD, spinal canal, and sacrum body semantic masks in 180 NAKO cases. We trained a default nnUNet model \cite{isensee_nnu-net_2021} to replicate this segmentation. This model and the Spinal Cord Toolbox \cite{de_leener_sct_2017} was used to segment our NAKO training data.

Adopting Graf et al.’s approach \cite{graf_denoising_2023}, we translated the T2w sagittal images into artifical CTs and used the Bonescreen SpineR tool (Bonescreen GmbH, Germany) based on Sekuboyina et. al. \cite{sekuboyina_btrfly_2018} to segment the artificial CTs. These translated annotations yielded nine vertebrae substructures segmentations (corpus, arcus vertebrae, spinous processus, and processus articulares inferiores, superiores and costales/transversus, the latter three divided into left and right). The translation process failed for 543 subjects, which were excluded from further usage.

These translation-based segmentation were added to the Streckenbach-based ones. When adding, we exclude voxels already segmented as a different structure. Next, we incorporated the spinal cord pixels. Finally, we filled holes between corpus and IVD regions and relabeled the transition pixels as endplates (see Figure \ref{fig:combine}).
This results in 14 spinal structures: ten for the vertebral substructures (including endplates), spinal canal, spinal cord, sacrum and intervertebral disc (IVD). This served as reference annotations for our training with NAKO data.

\begin{figure*}[!htb]
    \centering
    \includegraphics[width=0.8\textwidth]{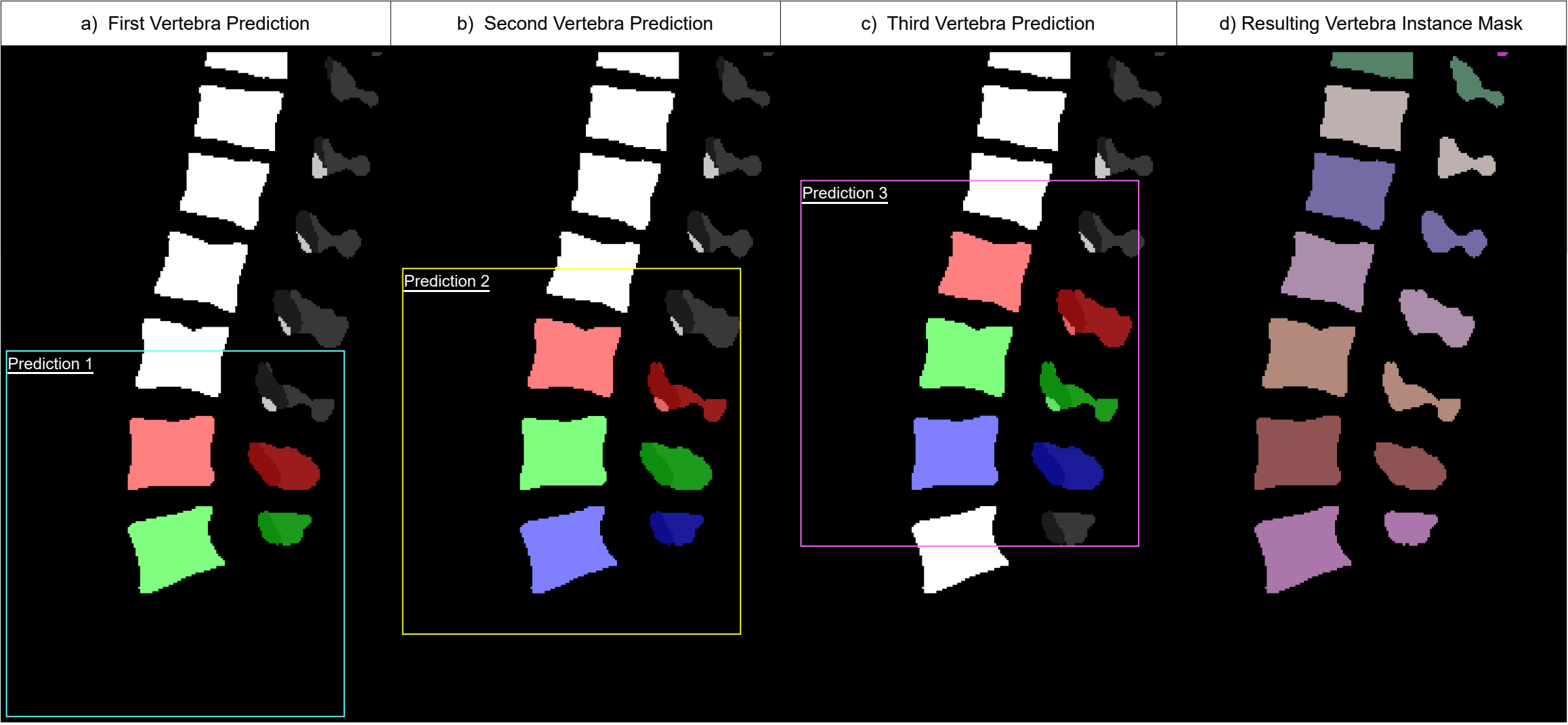}
    \caption{Given the semantic segmentation, we create cutouts of exact same size. Each of those cutouts (colored boxes) is fed into the instance model. a), b), and c) show the first three predictions predictions of a semantic input. The instance model always predicts the center vertebra of the cutout (green), as well as the one above (red) and below (blue), if visible. Therefore, assuming no erroneous behavior, we get three predictions for all inner vertebrae and two for the outer ones. For example, the second to last vertebra in the figure is predicted thrice, once in each of the three predictions (red, green, blue, from left to right). The combination of all cutout predictions are combined into a vertebra instance mask d), labeling each vertebra instance uniquely (different colors).}
    \label{fig:instance}
\end{figure*}

\subsection{Segmentation Approach}
Our approach operates in two phases (see Figure \ref{fig:workflow}). Initially, a semantic model segments the scan patch-wise into the 14 semantic labels. For this purpose, we employ a nnUNet architecture \cite{isensee_nnu-net_2021}. This yields the semantic mask.
The different instances cannot be trivially computed from this semantic mask, e.g. due to the fusion of vertebrae. Therefore, we utilize a sliding window approach with a second model to distinguish semantic labels into vertebra instances. To achieve this, we compute the center of mass positions for each vertebra corpus in the semantic mask through connected components analysis. Cutouts of fixed size (248, 304, 64) are created around these centers. For each cutout, the model predicts the three vertebrae around the cutout’s center. During this process, each vertebra appears in multiple cutouts (see Figure \ref{fig:instance}).
To reconcile different predictions, we calculate Dice scores between different predictions and sort them into triplets by their inter-dice agreement. The order in which the vertebra instance mask is finalized is based on the highest to lowest inter-dice agreements, guaranteeing least-consistent predictions are addressed last. Unlike previous approaches, this method ensures that we neither skip an entire vertebra nor merge two vertebrae into one. As the input to the instance model is the semantic mask, we do not use the image data during this process.
Finally, intervertebral discs (IVDs) and endplate structures from the semantic mask are added to the instance mask and given instance labels based on the nearest vertebra instance above.

For the detailed configurations used for training each model, see Appendix A.

\subsection{Post-Processing}
After semantic and instance segmentation, we fill holes in the result masks and zero out pixels in the instance mask that are zero in the semantic mask. Furthermore, we assign each connected component present in the semantic mask but absent in the instance mask to the instance with the highest number of neighboring voxels. This ensures consistency in foreground voxels between both masks.

\subsection{Experiments}
We used the nnUNet approach presented by van der Graaf et al. as a baseline \cite{van_der_graaf_lumbar_2023}. We compared its performance to our SPINEPS approach, training solely on the SPIDER dataset and evaluating on the SPIDER test split.

To demonstrate the effectiveness of the automated annotations, we used our approach trained only on the automated annotations of the NAKO training data and evaluated on the manually corrected NAKO test set. Additionally, we compared the performance with the automated annotations directly and a model trained on both NAKO training data and the SPIDER dataset.

As qualitative evaluation, we used our approach on the in-house out-of-distribution dataset, and manually reviewed the resulting segmentations.

\begin{figure*}[!t]
    \centering
    \includegraphics[width=0.8\textwidth]{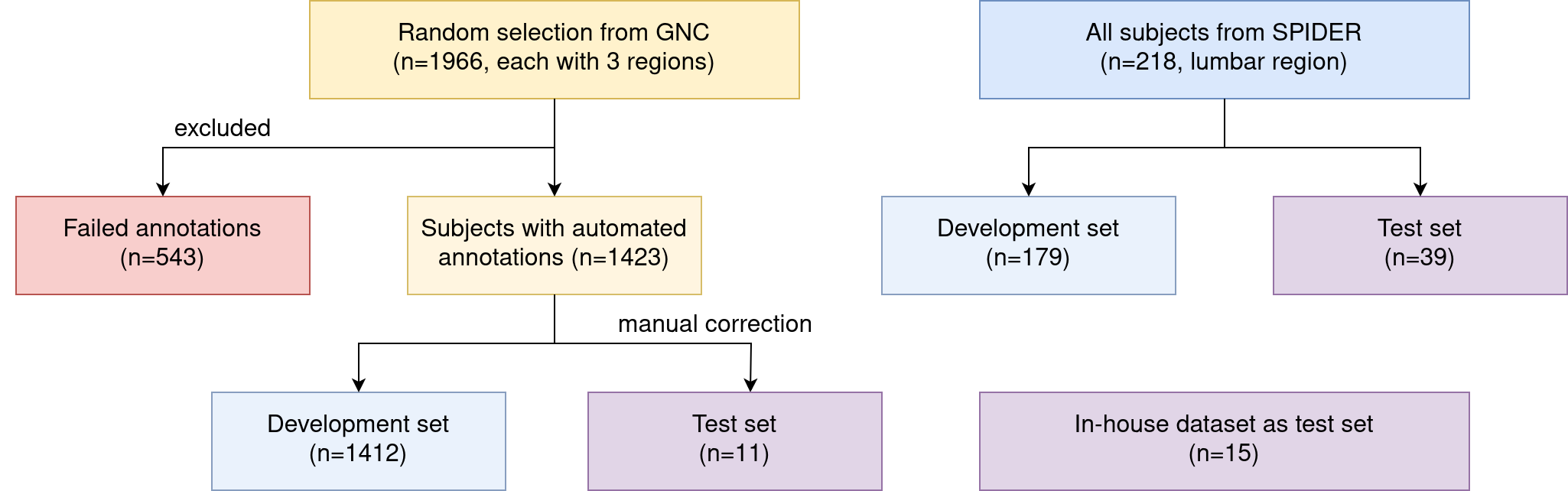}
    \caption{Flow diagram for subject exclusion from top to bottom for the different datasets. We only excluded subjects from the German National Cohort (GNC) because some of the automated annotation generation approaches failed for some subjects.}
    \label{fig:cohort}
\end{figure*}
\begin{table*}[!b]
\centering
\begin{tabular}{llccc}
\toprule
Structure        & Metric     & nnUNet 				Baseline & SPINEPS w/o post-processing & SPINEPS                                    \\ \midrule
\multicolumn{5}{c}{Global 				structure-wise}                                                                                                       \\ \midrule
Vertebra         & ↑ DSC      & 0.924 				± 0.025   & \textbf{0.938 }				± 0.022                     & 
\textbf{0.938 }				± 0.022 \\ 
IVD              & ↑ DSC      & 0.891 				± 0.036   & \textbf{0.904 }				± 0.033                     &   \textbf{0.904 }				± 0.033                                                   \\
Spinal 				Canal & ↑ DSC      & 0.922 				± 0.026   & \textbf{0.934 }				± 0.024                     & \textbf{0.934 }				± 0.024                                                      \\ \midrule
\multicolumn{5}{c}{Instance-wise}                                                                                                                   \\ \midrule
Vertebra         & ↑ DSC      & 0.907 				± 0.10    & 0.922 				± 0.081                      & \textbf{0.929 }				± 0.082                                   \\
                 & ↑ RQ       & 0.953 				± 0.12    & 0.984 				± 0.041                      & \textbf{0.992 }				± 0.03                                    \\
                 & ↑ SQ       & 0.856 				± 0.04    & 0.874 				± 0.031                      & \textbf{0.882 }				± 0.03                                    \\
                 & ↑ PQ       & 0.819 				± 0.11    & 0.859 				± 0.045                      & \textbf{0.875 }				± 0.038                                   \\
                 & ↓ 				ASSD & 0.533 				± 1.435   & 0.296 				± 0.356                      & \textbf{0.264 }				± 0.363                                   \\ \midrule
IVD              & ↑ DSC      & 0.873 				± 0.11    & \textbf{0.896 }				± 0.072                     & \textbf{0.896 }				± 0.072                                   \\
                 & ↑ RQ       & 0.951 				± 0.12    & 0.982 				± 0.053                      & \textbf{0.983 }				± 0.053                                   \\
                 & ↑ SQ       & 0.804 				± 0.06    & \textbf{0.824 }				± 0.054                     & \textbf{0.824 }				± 0.054                                   \\
                 & ↑ PQ       & 0.77 				± 0.12     & 0.808 				± 0.065                      & \textbf{0.81 	}			± 0.065                                    \\
                 & ↓ 				ASSD & 0.656 				± 2.46    & \textbf{0.277 }				± 0.216                     & \textbf{0.277 }				± 0.216                                  
\end{tabular}
\caption{The performance comparison between the nnUNet baseline adapted from van der Graaf et al. \cite{graf_denoising_2023} and our SPINEPS approach on the test split of the SPIDER dataset. To fairly compare how much our post-processing systems contribute, the metrics of SPINEPS without post-processing are also shown. We did not employ any post-processing for the semantic model, so metrics didn’t change there. Our approach outperforms the baseline in every metric, especially in the instance-wise metrics. The biggest difference can be seen in the instance-wise IVD ASSD metric, where our approach plus one standard deviation is still better than the average of the baseline Mean and standard deviations are reported. The arrows before the metric name indicate if smaller or higher values are better. We mark the best results in the comparison with an asterisk. DSC = Dice similarity coefficient, RQ = Recognition Quality, SQ  = Segmentation Quality, PQ = Panoptic Quality, ASSD = Average symmetric surface distance.}
\label{tab:baseline}
\end{table*}

\subsection{Statistical Analysis}
For evaluation, we employed the Dice similarity coefficient (DSC) and the average symmetric surface distance (ASSD), indicating average distances from segmented edges to reference annotations. Instance-wise metrics—recognition quality (RQ), segmentation quality (SQ), and panoptic quality (PQ), as described in \cite{kirillov_panoptic_2019}, calculated using panoptica \cite{kofler_panoptica_2023} —provided insights into instance prediction performance. Instances with an intersection over union (IoU) greater than or equal to 0.5 were considered true positives.
Statistical significance was determined using the Wilcoxon signed-rank test on Dice and recognition quality metrics, with p<0.05 indicating significance.

\section{Results}
\label{sec:experiments}

\subsection{Participants}
Table \ref{tab:demographics} presents the baseline demographic and clinical characteristics of the subjects. Out of the total 1966 subjects from the NAKO subset, we managed to get automated annotations for 1423 of them (mean age 53, 49\% female). We observed no disease-related pattern in our exclusion set, like strong scoliosis or hyper-intense spots. Eleven subjects from the 1423 were randomly selected as test split and manually corrected by experts (see Figure \ref{fig:cohort}). The public SPIDER dataset \cite{van_der_graaf_lumbar_2023} does not provide demographics aside from 63\% rate of female subjects. 39 subjects out of the 218 in the SPIDER dataset were randomly selected as a test split. The in-house dataset of 15 subjects was used for qualitative evaluation.

\begin{table*}[htb]
\centering
\begin{tabular}{llccc}
\toprule
Structure                & Metric     & Automated 				Annotations & SPINEPS 				(GNC) & SPINEPS 				(GNC + SPIDER) \\ \midrule
\multicolumn{5}{c}{Global 				structure-wise}                                                                      \\ \midrule
Vertebra                 & ↑ DSC      & 0.855 				± 0.033         & 0.90 				± 0.029  & \textbf{0.92 	}			± 0.029          \\
IVD                      & ↑ DSC      & 0.942 				± 0.022         & 0.960 				± 0.013 & \textbf{0.967 }				± 0.01          \\
Spinal 				Canal         & ↑ DSC      & 0.928 				± 0.026         & 0.947 				± 0.023 & \textbf{0.958 }				± 0.064         \\
Spinal 				Cord          & ↑ DSC      & 0.932 				± 0.038         & 0.959 				± 0.021 & \textbf{0.966 }				± 0.014         \\ \midrule
\multicolumn{5}{c}{Vertebra 				substructures}                                                                     \\ \midrule
Arcus 				Vertebra       & ↑ DSC      & 0.79 				± 0.062          & 0.853 				± 0.069 & \textbf{0.871 }				± 0.069         \\
                         & ↓ 				ASSD & 0.783 				± 0.439         & 0.38 				± 0.286  & \textbf{0.303 }				± 0.261         \\ \midrule
Spinosus 				process     & ↑ DSC      & 0.747 				± 0.066         & 0.811 				± 0.064 & \textbf{0.838 }				± 0.062         \\
                         & ↓ 				ASSD & 0.73 				± 0.44           & 0.498 				± 0.322 & \textbf{0.356 }				± 0.272         \\ \midrule
Articularis 				inferior & ↑DSC       & 0.69 				± 0.093          & 0.763 				± 0.09  & \textbf{0.79 	}			± 0.092          \\
                         & ↓ 				ASSD & 0.8 				± 0.51            & 0.568 				± 0.4   & \textbf{0.46 	}			± 0.367          \\ \midrule
Articularis 				superior & ↑ DSC      & 0.634 				± 0.096         & 0.736 				± 0.974 & \textbf{0.762 }				± 0.107         \\
                         & ↓ 				ASSD & 1.27 				± 0.693          & 0.72 				± 0.453  & \textbf{0.643 }				± 0.593         \\ \midrule
Costal 				process       & ↑ DSC      & 0.541 				± 0.1           & 0.631 				± 0.102 & \textbf{0.683 }				± 0.1           \\
                         & ↓ 				ASSD & 1.95 				± 0.79           & 1.58 				± 1.2    & \textbf{1.14} 				± 0.613          \\ \midrule
Vertebra 				corpus      & ↑ DSC      & 0.904 				± 0.02          & 0.942 				± 0.015 & \textbf{0.96 	}			± 0.014          \\
                         & ↓ 				ASSD & 0.756 				± 0.295         & 0.341 				± 0.126 & \textbf{0.195 }				± 0.082        
                         \\ \bottomrule
\end{tabular}
\caption{The performance comparison between the automated annotations, SPINEPS trained only on the GNC and SPINEPS trained with both the GNC and the SPIDER dataset. Mean and standard deviations are reported. Evaluation is done on the manually corrected test split from the GNC. Our approach outperforms the automated annotations, yet incorporating the manually annotated SPIDER dataset improves each metric further. The arrows before the metric name indicate if smaller or higher values are better. We mark the best results in the comparison with an asterisk. IVD = Intervertebral Disc, DSC = Dice similarity coefficient, ASSD = Average symmetric surface distance, GNC = German National Cohort.}
\label{tab:gnc}
\end{table*}

\subsection{Performance}
\begin{figure*}[!htb]
    \centering
    \includegraphics[width=0.8\textwidth]{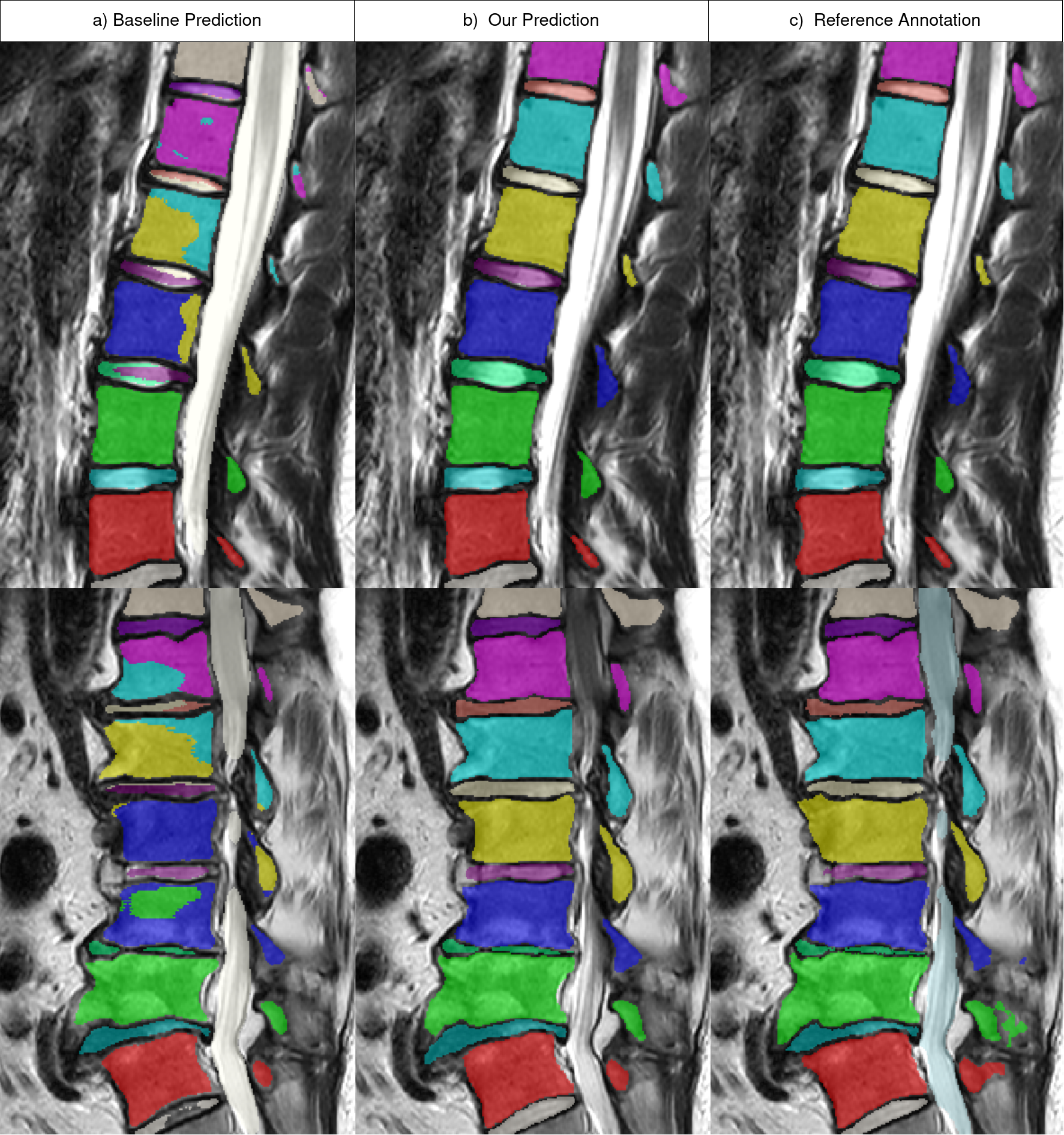}
    \caption{Example subjects where the baseline (a) produces a typical found error: mixing different instance labels. Our approach (b) is very close to the reference annotation (c). This type of error the baseline made did not occur with our approach.}
    \label{fig:spider}
\end{figure*}

To showcase the performance difference for our SPINEPS approach, we only used the SPIDER dataset for training and evaluated on the SPIDER test set for the first experiment. Table \ref{tab:baseline} compares the baseline nnUNet to our two-phase approach. SPINEPS outperforms the baseline across all metrics and structures (all dice p-values <0.0001). Even without the proposed post-processing techniques, our two-phase approach outperforms the baseline. Contrary to the baseline, our model does not produce global instance segmentation errors (see Figure \ref{fig:spider}).

For the NAKO dataset, we directly compared the automated annotations, a model trained on those annotations, and a model additionally trained with the SPIDER dataset training split (see Table \ref{tab:NAKO}). Using a model trained on the automated annotations from NAKO outperformed the automated annotations. Adding the SPIDER dataset for training increased the performance as well (all dice p-values $<0.0001$). We also evaluated the individual regions (cervical, thoracic, lumbar) using our approach (see Appendix B: Performance by Region).

\begin{figure*}[!htb]
    \centering
    \includegraphics[width=0.8\textwidth]{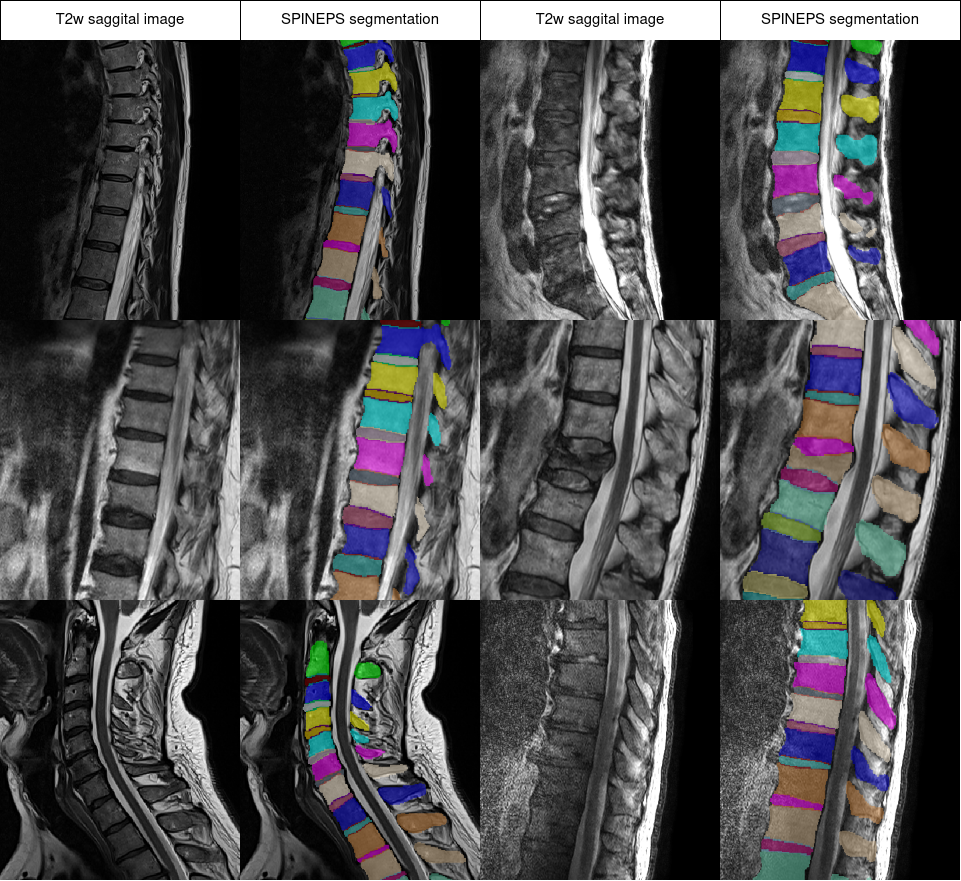}
    \caption{Some MRIs from our in-house dataset and the instance masks created by our approach (SPINEPS) trained on the automated annotations of the German National Cohort. It shows the robustness towards out-of-distribution scans, derived from different MRI devices, field strengths, spatial resolutions, and with pathologies. Ground truth annotations do not exist to compare our prediction to. Qualitatively, we observe that the approach adapts well to unknown devices and aberrant cases, producing no global segmentation errors.}
    \label{fig:ood}
\end{figure*}

We used our best SPINEPS setup on our in-house dataset containing out-of-distribution samples. In this case, acquisition of different scanners, field strengths, spatial resolutions and containing pathologies. After a qualitative review, we did not observe any major segmentation quality differences compared to the test samples from the NAKO (see Figure \ref{fig:ood}).

\subsection{Related Comparison}
To put our work in perspective, we picked two state-of-the-art approaches in spine segmentation and compared our results. We show that we achieved similar or better metric scores while showing clear advantages, e.g. being able to segment vertebra substructures including posterior elements and designed to work not only on the lumbar, but the whole spine (see Table \ref{tab:related}).

\begin{table*}[htb]
    \centering
    \begin{tabular}{rx{2.5cm}x{2.5cm}x{2.5cm}x{2.5cm}}
    \toprule
    Attribute                           & SPINEPS 				(ours)   & SPIDER 				nnUNet baseline (18) & Li 				et al. (11) & Sáenz-Gamboa 				et. al. (9)                 \\ \midrule
    Input MRI                               & T2w 				sagittal & T2w 				sagittal            & T2w 				axial  & T1w 				and T2w sagittal (requires both) \\
    Dimensionality                      & 3D                   & 3D                              & 2D 				slices      & 2D 				slices                                \\
    Outputs 				semantic mask           & Yes                  & No                              & Yes                & Yes                                          \\
    Outputs 				instance mask           & Yes                  & Yes                             & No                 & No                                           \\
    Lumbar 				region                   & Yes                  & Yes                             & Yes                & Yes                                          \\
    Thoracic 				region                 & Yes                  & No                              & No                 & No                                           \\
    Cervical 				region                 & Yes                  & No                              & No                 & No                                           \\
    Vertebra 				Corpus DSC ↑           & 0.94                 & N/A                             & 0.98               & N/A                                          \\
    Vertebra 				Corpus ASSD ↓          & 0.34                 & N/A                             & 2.32               & N/A                                          \\
    Global 				Vertebra DSC ↑           & 0.92                 & 0.92                            & N/A                & 0.93                                         \\
    Global 				IVD DSC ↑                & 0.96                 & 0.89                            & N/A                & 0.94                                         \\
    Global 				spinal canal DSC ↑       & 0.95                 & 0.92                            & N/A                & 0.88                                         \\
    Includes 				posterior elements     & Yes                  & Yes                             & No                 & Yes                                          \\
    Includes 				vertebra substructures & Yes                  & No                              & No                 & No                                           \\
    Publicly 				available              & Yes                  & Yes                             & No                 & No                                          \\ \bottomrule
    \end{tabular}
    \caption{Comparison of our SPINEPS approach compared to two approaches trying to solve similar problems. Values are only stated if they were reported in the respective paper and we are able to report the same category. The metric values are not scientifically comparable, yet give a rough estimate on how our approach compares to some state-of-the-art approaches. Thus we don’t mark best values in this table. Although Li et al. \cite{li_automatic_2021} has a higher DSC for the vertebra corpus structure, their ASSD is six times higher than ours. Sáenz-Gamboa et. al. \cite{saenz-gamboa_automatic_2023} achieved a higher global vertebra DSC, but we outperform them on every other structure, without the need of T1w images, and not restricted to the lumbar region. The arrows after the metric name indicate if smaller or higher values are better. IVD = Intervertebral Disc, DSC = Dice similarity coefficient, ASSD = Average symmetric surface distance}
    \label{tab:related}
\end{table*}

\section{Discussion}
\label{sec:discussion}

This work presents the first publicly available algorithm for whole spine segmentation in sagittal T2w MR images. So far, only models that either segment only few structures (e.g. vertebral body) or focus on one region (e.g. lumbar spine) were available. We demonstrated that our two-phase approach yields improved semantic and instance segmentation capabilities. Training a model on combined automated annotations, partially yielded from a MR to CT translation approach, results in good performance and generalization.

Our approach is trained on a large cohort of MRI data available from the German National Cohort (NAKO). Such a segmentation technique enables further studies and essential analysis, such as deriving normative values, as shown by Streckenbach et al. (18). However, in that study, the proposed model only segments the vertebra corpus, IVD, and spinal canal region semantically. Further algorithms are required to derive the instance masks from their semantic outputs, rendering downstream tasks more challenging.

We show samples where the baseline nnUNet model produces global segmentation errors, as the different labels are of the same semantic structure and look similar to the model. With our approach, we avoid this by first segmenting semantically and using the proposed sliding window technique on fixed cutout sizes.
Using our two-phase approach for semantic and instance segmentation, we outperform a baseline on a public dataset. The increases in performance regarding average symmetrical surface distance (ASSD) and the lower standard deviations suggest a higher robustness of our technique. Additionally, contrary to the baseline, our approach yields both a semantic and an instance mask. This is crucial to allow for downstream tasks such as fracture detection that are interested in the semantic structures as well as the localization of individual instances. Although our instance map only contains vertebrae, IVDs, and endplates, all other semantically segmented structures (spinal canal, spinal cord, sacrum) have only one instance by nature, thus the instance information can be derived from the semantic mask directly.

Training on the automated annotations generated mainly from a translation approach, we show a DSC score of 0.90 for vertebrae, 0.960 for IVDs, and 0.947 for the spinal canal on 11 randomly chosen, manually corrected subjects. Adding manual annotations like those from the SPIDER dataset improves these metrics further, the global vertebra DSC increases from 0.90 to 0.92. We have shown and compared some state-of-the-art approaches in the area of spine segmentation with our approach. Thus, we consider our approach trained on automated annotations to be highly competitive, achieving similar or better results on the whole spine, without the need to manually segment the whole training data.

Given the required training data, our approach can be adapted to any image modality for spine segmentation. While the model trained on the SPIDER dataset works for both T1w and T2w sagittal views, our model trained on the automated annotations of NAKO samples can currently only be used for T2w sagittal views. Nevertheless, our instance phase works on the semantic mask output of the first model without image data and, thus, works for any given semantic segmentation of vertebrae, regardless of origin. This makes our approach more feasible to adapt to other MRI modalities.

There are limitations to our study. The population of the NAKO represents an average German, healthy population. Thus, in comparison to a typical hospital dataset, pathologies may be under-represented, and we cannot guarantee the same performance on image data such as postoperative MRI.
The samples from the NAKO often contain only a few slices and have a slice thickness of 3.3mm. We noted that our model occasionally encounters difficulty in accurately segmenting the outermost voxels along the left/right dimension. This leads to pixel omissions in the processus costalis/transversus structures, making it the least proficient substructure in our model's performance. Lastly, our model is unable to segment C1 vertebrae, as none of our training data had that particular vertebra segmented.

\raggedbottom
\section{Conclusion}
We presented SPINEPS, a two-phase approach to semantic and instance segmentation. It is superior to a nnUNet baseline, while additionally providing instance segmentation capabilities. We showed our approach to generalize well using only automatic annotations, partially derived from an MR to CT translation approach. Finally, we presented a whole spine model to accurately and robustly segment 14 spinal structures in T2w sagittal scans in cervical, thoracic, and lumbar regions, both semantically and instance-wise. We made the models and approach publicly available1, thus enabling segmentations for the German National Cohort and other datasets to be used in further spine studies.





\section*{Acknowledgement}

This project was conducted with data from the German National Cohort
(NAKO) (www.nako.de). The NAKO is funded by the Federal Ministry of
Education and Research (BMBF) [project funding reference numbers:
01ER1301A/B/C, 01ER1511D and 01ER1801A/B/C/D], federal states of
Germany and the Helmholtz Association, the participating universities and
the institutes of the Leibniz Association.
We thank all participants who took part in the NAKO study and the staff of
this research initiative.

\clearpage
{
    \small
    \bibliographystyle{unsrt}
    \bibliography{references}
}

\clearpage
\clearpage
\onecolumn
\setcounter{section}{41}
\section{Supplemental Material}
\label{sec:supplement}

\subsection{Appendix A: Training Procedure}
We prepared the data by reorienting each image consistently and re-sampling them to the same resolution. For the experiments using only SPIDER, we used the resolution suggested by the nnUNet framework \cite{isensee_nnu-net_2021}. For the NAKO-based model, we used a resolution of (0.75, 0.75, 1.65), where the third dimension is the left/right axis. When training with the NAKO images, we cropped them slightly in each dimension. This reduced the errors made by the automatic annotations process. Additionally, we pre-processed the T2w images by applying the N4 Bias field correction algorithm \cite{tustison_n4itk_2010}.
As the annotations in the SPIDER dataset \cite{van_der_graaf_lumbar_2023} only consist of instance labels, we used an in-house available segmentation model based on \cite{sekuboyina_btrfly_2018} to create three subregion labels for the vertebrae but only utilized this for training purposes.

We utilized the widely used nnUNet \cite{isensee_nnu-net_2021} as semantic model on the combined annotations. As the annotations are not rotation-invariant, we disabled the Mirroring Augmentation. Besides setting the patch size to (256, 256, 64), we used the automatically calculated parameters from the nnUNet framework. We trained for 1000 epochs, a batch size of 2, and with 3-fold cross-validation. The baseline was trained identically, but on the instance-labels, not the semantic ones. For inference, we let all 3 folds run a prediction and average their outputs.

For the instance phase, we chose a cutout size of (248, 304, 64) to always contain three complete vertebrae. We trained an Unet3D model on these cutouts to segment the three vertebrae around the center with the labels 1 (above), 2 (center), and 3 (below). We used a batch size of 2, a learning rate of 1e-4 that decreases linearly each epoch to 1e-6, and trained for 300 epochs. For augmentations, we used random scaling in the range [-0.2,0.2], Random Erosion, Random Down- and then Upsampling, Random Labeldrop, each with a 10\% chance of occurring during training. With a 25\% chance, we adopted Random Vertical Crop. We used Horizontal Flip to as augmentation, practically doubling the training data.

\clearpage
\subsection{Appendix B: Performance by Region}
We compared the performance of our best model for each region. Each subject in the German National Cohort (NAKO) has scans for the cervical, thoracic, and lumbar spine. Table 4 shows the performance of the model trained both on the NAKO and SPIDER dataset on the NAKO test set, separated for the different regions. Although the global vertebra dice score is best at the lumbar region, our trained model overall performs best on the thoracic region, followed by the lumbar and then the cervical region. We hypothesize the cervical regions to be worst because the substructures of the vertebrae are very difficult to distinguish in T2w sagittal TSE images there. Even the CT segmentations that were used were performing worse in the upper cervical region, further hinting at this underlying problem.

\begin{table*}[htb]
    \centering
    \begin{tabular}{llccc}
    \toprule
    Structure                & Metric     & Cervical 				region & Thoracic 				region & Lumbar 				region  \\ \midrule
    \multicolumn{5}{c}{Global 				structure-wise}                                                          \\ \midrule
    Vertebra                 & ↑ DSC      & 0.889 				± 0.018   & 0.929 				± 0.014   & \textbf{0.937 }				± 0.028 \\
    IVD                      & ↑ DSC      & 0.962 				± 0.008   & \textbf{0.97 	}			± 0.003   & 0.969 				± 0.011  \\
    Spinal 				Canal         & ↑ DSC      & 0.96 				± 0.006    & \textbf{0.961 }				± 0.008  & 0.952 				± 0.02   \\
    Spinal 				Cord          & ↑ DSC      & \textbf{0.975 }				± 0.006  & 0.97 				± 0.008    & 0.954 				± 0.015  \\ \midrule
    \multicolumn{5}{c}{Vertebra 				substructures}                                                         \\ \midrule 
    Arcus 				Vertebra       & ↑ DSC      & 0.841 				± 0.05    & \textbf{0.90 	}			± 0.03    & 0.87 				± 0.091   \\
                             & ↓ 				ASSD & 0.394 				± 0.22    & \textbf{0.202 }				± 0.077  & 0.314 				± 0.36   \\ \midrule
    Spinosus 				process     & ↑ DSC      & 0.807 				± 0.035   & \textbf{0.857 }				± 0.034  & 0.85 				± 0.086   \\
                             & ↓ 				ASSD & 0.498 				± 0.32    & \textbf{0.265 }				± 0.085  & 0.309 				± 0.28   \\ \midrule
    Articularis 				inferior & ↑ DSC      & 0.714 				± 0.088   & \textbf{0.827 }				± 0.029  & 0.823 				± 0.091  \\
                             & ↓ 				ASSD & 0.701 				± 0.467   & \textbf{0.296 }				± 0.093  & 0.383 				± 0.298  \\ \midrule
    Articularis 				superior & ↑ DSC      & 0.683 				± 0.1     & \textbf{0.813 }				± 0.053  & 0.787 				± 0.11   \\
                             & ↓ 				ASSD & 0.881 				± 0.522   & \textbf{0.352 }				± 0.151  & 0.691 				± 0.77   \\ \midrule
    
    Costal 				process       & ↑ DSC      & 0.641 				± 0.079   & \textbf{0.71 	}			± 0.966   & 0.701 				± 0.107  \\
                             & ↓ 				ASSD & 1.546 				± 0.511   & 1.19 				± 0.623    & \textbf{0.742 }				± 0.397 \\ \midrule
    Vertebra 				corpus      & ↑ DSC      & 0.95 				± 0.013    & 0.964 				± 0.01    & \textbf{0.966 }				± 0.014 \\
                             & ↓ 				ASSD & 0.188 				± 0.087   & \textbf{0.191 }				± 0.058  & 0.2 				± 0.093   \\ \bottomrule
    \end{tabular}
    \caption{The performance of the SPINEPS model trained on both the automated annotations and the SPIDER dataset, evaluated on the GNC test set. We evaluated each region (cervical, thoracic, lumbar) individually for comparison. As each GNC subject has scans for each region, the total number of data points are equal, and thus, comparable. Mean and standard deviations are reported. Evaluation is always done on the manually corrected test split from the GNC. Although the global vertebra DSC is best at the lumbar region, our model trained on both datasets overall performs best at the thoracic region, closely followed by the lumbar region, and then cervical region. The arrows before the metric name indicate if smaller or higher values are better. We mark the best results in the comparison with an asterisk. IVD = Intervertebral Disc, DSC = Dice similarity coefficient, ASSD = Average symmetric surface distance, GNC = German National Cohort.}
    \label{tab:region}
\end{table*}

\end{document}